\documentclass[12pt]{article}
\usepackage{amsmath,amssymb,psfig}

\newtheorem{theorem}{\large\bf Theorem}

\newcommand{\RR}{{\mathbb R}}          %%%%% Reali
\newcommand{\be}{\begin{equation}}
\newcommand{\ee}{\end{equation}}
\newcommand{\bqa}{\begin{eqnarray}}
\newcommand{\eqa}{\end{eqnarray}}
\newcommand{\ba}{\begin{array}}
\newcommand{\ea}{\end{array}}

\newcommand{\no}{\nonumber}

\newcommand{\al}{\alpha}

\newcommand{\ga}{\gamma}
\newcommand{\de}{\delta}
\newcommand{\ep}{\epsilon}

\newcommand{\la}{\lambda}

\newcommand{\ch}{\chi}

\newcommand{\Ph}{\Phi}
\newcommand{\si}{\sigma}

\newcommand{\cA}{{\cal A}}
\newcommand{\cB}{{\cal B}}
\newcommand{\cC}{{\cal C}}
\newcommand{\cL}{{\cal L}}

\begin{document}

\centerline{\large \bf
Supersymmetric Analysis of a Simplified }
\centerline{\large \bf
Two Dimensional Anderson Model at Small Disorder}
\vskip 2cm

\centerline{J. Bellissard$^1$, J. Magnen$^2$ and V. Rivasseau$^3$}

\centerline{1. Universit{\'e} Paul Sabatier et Institut Universitaire de France}
\centerline{118 Route de Narbonne, 31062 Toulouse Cedex 04}
\medskip

\centerline{2. Centre de Physique Th{\'e}orique, CNRS UMR 7644}
\centerline{Ecole Polytechnique, 91128 Palaiseau Cedex, FRANCE}
\medskip

\centerline{3. Laboratoire de Physique Th{\'e}orique, CNRS UPR 7644}
\centerline{Universit\'e d Orsay, 91405 Orsay, FRANCE}

\vskip 1cm
\medskip
\noindent{\bf Abstract}
This work proposes a very simple random matrix model, the Flip Matrix Model, 
liable to approximate the behavior of a two dimensional electron in a weak 
random potential. Its construction is based on a phase space analysis, a 
suitable discretization and a simplification of the true model. The density of 
states of this model is investigated using the supersymmetric method and shown 
to be given, in the limit of large size of the matrix by the usual Wigner's 
semi-circle law.

%%%%%%%%%%%%%%%%%%%%%%%%%%%%%%%%%%%%%%%%%%%%%%%%%%%%%%%%%%%%%%%%%%%%%%%%%%%%%%%%
\section{Introduction}
\label{susy.sec-intro}

\noindent This paper is devoted to the rigorous study of a random matrix model 
that is liable to give a good approximation for the density of states (DOS) of 
the two dimensional Anderson model in the weak disorder regime and in the 
vicinity of the Fermi level. The main result is that the DOS of this random 
matrix model converges to the usual Wigner's semi-circle law in the small 
coupling limit. The Anderson model of an electron in a random potential 
corresponds to the Hamiltonian

%%%%%%%%%%%%%%%
$$  
H \;=\;  (- \Delta  + \la V(x))
$$
%%%%%%%%%%%%%%%

\noindent acting on the Hilbert space $L^2 (\RR^2)$, where $\Delta$ is the usual 
Laplacian and $V$ is a real Gaussian process on $\RR^2$ with short range 
correlations. 

\noindent This model was initially proposed to describe the electron motion
in doped semi-conductors at low temperature or in normal disordered
metals. It is now the central model for the theory of eletronic transport
and wave propagation in disordered systems \cite{EFSHK}. It was conjectured by 
Anderson as soon as 1958 \cite{And58} that such a model exhibits a localized 
phase in which the electrons are trapped by the defects. In 1979 it was argued 
that this model has a phase transition in dimensions three or more between the 
localized phase and an extended one \cite{gang4}.

\vspace{.1cm}

\noindent The localized phase is now well under control. In one dimension, 
localization was rigorously established for any disorder at the end of the 
seventies \cite{GMP,KuSo}. Later localization was established in any dimension 
at strong disorder or for energies out of the conduction bands \cite{FMSS}. A 
simplified and more efficient method to get this result was given in 
\cite{AiMo}.

\vspace{.1cm} 

\noindent In contrast the weak disorder regime is still poorly understood. In 
two dimensions it has been argued \cite{gang4,Bergmann} and numerically 
established \cite{Kra} that localization persists at arbitrarily small disorder, 
with a localization length diverging like $O(e^{c/\la^{2}})$. In dimension 
three, numerical simulations confirm the existence of the Anderson transition 
\cite{Kra}, leading to an extended phase. In addition, analytical results 
\cite{Efet83} and other numerical 
calculations show that the level spacing distribution follows the Wigner-Dyson 
distribution for random matrix theory (RMT) \cite{Kra2}. This gave the 
motivation 
for a description of mesoscopic system in terms of RMT \cite{AltShk}. This 
method has been very successful when compared to experiments and was the source 
of developments of supersymmetric methods \cite{Efet,Mir} in solid state 
physics. But on the rigorous level it is still a mathematical challenge 
up to now to prove even regularity of the DOS (e.g. real analyticity in energy 
in a certain interval) at arbitrarily weak coupling in dimensions two and three.

\noindent Using a phase-space analysis inspired by the renormalization group 
method around Fermi surfaces in condensed matter \cite{BG,FT,FMRT}, G. Poirot 
and coauthors \cite{MPR1,Po} have established that the effective Hamiltonian 
near the Fermi level is given indeed by a random matrix model. In two dimensions 
this random matrix model is similar to the GUE, but contains an extra discrete 
symmetry. Unfortunately in three dimensions and beyond, this extra symmetry 
becomes continuous, and produces more complicated correlations between matrix 
elements \cite{MPR1}. Recently a band matrix model in three dimensions has been 
proposed and its DOS studied through the supersymmetric method \cite{DPS}. In 
this note, motivated by these works, the two dimensional case for the Anderson 
model is reconsidered. A simplified model is constructed on the basis of the 
phase space analysis. It is a matrix model, although not one of the regular 
ensembles. We analyze this model rigorously through the supersymmetric method, 
and proves that its DOS converges to the Wigner semi-circle law as the matrix 
gets large.

%%%%%%%%%%%%%%%%%%%%%%%%%%%%%%%%%%%%%%%%%%%%%%%%%%%%%%%%%%%%%%%%%%%%%%%%%%%
\section{The Flip Matrix Model}
\label{susy.sec-flipmatrix}

\noindent Average spectral properties can be studied through the averaged 
Green's functions of the model. For some suitable choice of units, these 
averaged Green's functions are defined as

%%%%%%%%%%%%%%%
$$  
G_{\pm }(y-x) \;=\;
 \lim _{\ep \to 0^{\pm}} 
  \int d\mu(V) 
   \langle x| (-\Delta - E+\imath \epsilon + \la V )^{-1}|y \rangle
$$
%%%%%%%%%%%%%%%

\noindent where $d\mu(V)$ is the disorder distribution, $\la$ is a coupling 
constant controlling the disorder strength, and $E$ is complex with small 
imaginary part $\ep$. The DOS is defined by the imaginary part of the averaged 
retarded Green's 
function :

%%%%%%%%%%%%%%%
$$ 
\nu (E) \;=\;
 \frac{1}{\pi}\;
  {\rm Im} G_{+}(0)
$$
%%%%%%%%%%%%%%%

\noindent For a given non-zero $E$ in the conduction band (e.g. for $\vert E 
\vert $ 
neither too large nor too small), the free Green's function $(p^2 - E)^{-1}$ (we 
simply write $-\Delta =p^{2}$) is singular on a surface, which in two dimensions
is a curve. Two regimes have been clearly identified:

%%%%%%%%%%%%%%%
\begin{enumerate}
\item for $\vert p^2 -E \vert \ge 0(1) \la ^2$, the random potential $\la V$
is indeed a weak perturbation of the free Green function. It is statistically 
unlikely to develop an eigenvalue of the same size than $p^2-E$, making the 
denominator of the Green's function singular. So this regime can be controlled 
by 
the usual techniques of multiscale cluster and Mayer expansion, with some 
large/small field analysis. Typical $V$'s belong to the small field regions; 
exceptional $V$'s in large field regions are controlled through some type of 
Tchebycheff inequalities, and give rise to small probabilistic factors. These 
factors can then pay for the necessary complex contour deformations and rough 
bounds
that desingularize the denominator of the Green's function in these regions 
\cite{Po}.

\item for $\vert p^2 -E \vert \le 0(1) \la ^2$, the interesting region, a non 
trivial imaginary part generated by the potential should stop the 
renormalization 
group analysis. This is easily seen at second order perturbation theory in 
$\la$: an 
imaginary term proportional to $i\la^2$ appears. Hence $G_{\pm}$ should decay 
exponentially with a length scale proportional to $\la^{-2}$,  and the DOS 
should be 
regular. The problem is to justify non pertubatively this well-known theoretical 
argument. 
\end{enumerate}
%%%%%%%%%%%%%%%

\noindent For the moment the main result in this direction is \cite{MPR2}:

%%%%%%%%%%%%%%%
\begin{theorem}
There exists some finite $\eta >0$ such that for $\ep = \la^{2+\eta}$
the averaged Green's function $< G_{+\ep} (x,y)>_{V}$ decays at large
$\vert x - y \vert $ with a
rate $\tau(\la)$ which is independent of $\eta$, and inverse of $\la^{2}$.
\end{theorem}
%%%%%%%%%%%%%%%

\noindent This result was obtained through some difficult non-perturbative {\em 
Ward identity} which uses the particular momentum conservation laws in two 
dimensions. It is non-trivial, since the  decay rate in $\la^{-2}$, as expected 
from perturbation theory, does not depend on the $\eta $ cutoff. This result 
seems difficult to extend to fully perform the limit $\ep \to 0$ (this would in 
fact be implied by a similar theorem but with $\eta = 1$).

\noindent To complete the proof of exponential decay of the averaged Green's 
function and of the regularity of the averaged DOS (without $\la^{2+\eta}$ 
cutoff), the supersymmetric method \cite{Efet,Mir} may be more convenient.
Indeed  the $\vert p^2 -E \vert \le 0(1) \la ^2$ regime of the 2D Anderson model
is no longer perturbative in the naive sense, since the potential combines 
non-trivially with the deterministic part. Supersymmetry seems the right tool to 
control this phenomenon since in this formalism it corresponds simply to contour 
deformation to a non-trivial saddle point that generates Wigner's law and stops 
the RG flow. The functional integral away from the saddle point can be 
controlled non-perturbatively by a small/large field analysis, which is 
compatible with standard weak-coupling cluster expansions \cite{DPS}.

\noindent The {\em Flip Matrix Model} studied in this paper incorporates 
four simplifications with respect to the true Anderson 2 dimensional model:

\begin{itemize}
  \item The regime $\vert p^2 -E \vert \ge 0(1) \la ^2$ has been completely 
  removed from the model. This is justified since this regime being fully
  perturbative can be added later in the style of \cite{Po}. So we restrict 
  our Hilbert space to functions supported in momentum space by a tiny shell 
  around the circle $p^2 =E$ (or $2-\cos p_1 - \cos p_2 =E$ for a square 
  lattice version of the model).

  \item The space has been reduced to a single cube of side size $\la^{-2}$. 
  This is in agreement with the idea of a contour translation generating a 
  $i \la^2$ part in the denominator of the Green's function. Then the full 
  model should correspond to weakly coupled such cubes. Once the single 
  cube-model has been understood, the full model should follow from a 
  cluster expansion.
  
  \item In this cube the momentum shell has been divided into $0(\la^{-2})$ 
  cells, called sectors, and the corresponding infinite dimensional Hilbert 
  space has been replaced by a finite dimensional one, so that the {\em random
  operator} corresponding to $V$ is replaced by a random matrix. This
  discretization step is justified since the phase space analysis is now fine
  enough, so that each random operator being approximately constant, it is
  legitimate to replace it by a single random coefficient.

  \item The probability law for the random coefficients of the matrix has 
  been computed from the 2D momentum conservation rule, with a last 
  simplification. As well known the {\em rhombus rule} of 2D momentum 
conservation 
  for four vectors of equal length slightly weakens for almost degenerate 
  {\em flat} rhombuses and this degeneracy requires an anisotropic slicing 
  of sectors and of the cube into parallelepipeds \cite{FMRT,Po,MPR2}. 
  We forget this complication to stick with simpler isotropic sectors, 
  so we use a simplified momentum conservation rule. Normally introducing
  anisotropic analysis is only a (painful) complication of the model 
  that gets rid of the difficulties associated to almost degenerate rhombuses.
\end{itemize}

\noindent Then the subsequent model would be almost the GUE if not for the 
rhombus 
{\em flip}. This flip creates a new symmetry in the random matrix, so that the 
supersymmetric analysis of this model requires essentially twice as many 
variables 
as for the GUE. The full proof of regularity of the DOS for the 2D Anderson 
model 
is therefore now reduced to the hard task of fitting together all pieces of the 
puzzle by removing our four approximations to go from the Flip Model to the true 
2D 
Anderson Model.

%%%%%%%%%%%%%%%%%%%%%%%%%%%%%%%%%%%%%%%%%%%%%%%%%%%%%%%%%%%%%%%%%%%%%%%%%%%
\section{The Flip Matrix Model: notations}
\label{susy.sec-flipnotation}

\noindent Let $N$ be a large number parametrizing the number of sectors. To 
model the $2D$ Anderson model in a large spatial cell of side size $O(\la^{-2})$ 
,we have divided as indicated above the momentum shell of width $O(\la^2)$ 
around $p^2=1$ into $2N= O(\la^{-2})$ isotropic cells. Since the initial $V(x)$ 
was real, its Fourier transform $V(p)$  (the hat on Fourier transforms is 
removed for simplicity) is complex and obeys 

%%%%%%%%%%%%%%%
\be
\label{susy.eq-hermitian}
 V(-p) = \bar V(p) \ . 
\ee
%%%%%%%%%%%%%%%

\noindent Since $V(x)$ is random independent identically distributed
in direct space, $V(p)$ is Gaussian white noise, so that

%%%%%%%%%%%%%%%
$$  
\langle V (p) \bar V (q) \rangle \;=\;
 \de (p-q)  \ .
$$
%%%%%%%%%%%%%%%

\noindent In order to implement the momentum conservation constraint at the 
vertex the "rhombus rule" will be simplified by forgetting the degeneracy of the 
rhombus near collapse. Discretization of  $V(p)$ follows from it being 
considered as a random matrix between the cell indices. Collision with the 
initial potential $V(p)$ transforms a state with momentum $r$ into a state with 
momentum $r+p$. Therefore the discretized matrix corresponds to $V$ with 
matrix elements $V(\ga)$ where $\ga$ is some discretized version of $p$. Its 
coefficients are

%%%%%%%%%%%%%%%
$$   
 V_{\al ,\beta}
$$
%%%%%%%%%%%%%%%

\noindent where $\al$ and $\beta $ run over the set of sectors and $ V_{\al 
,\beta}$ transforms a state with momentum sector $\al$ into a state with 
momentum sector $\beta$ if and only if $\ga \simeq \beta - \al$. 
Solving this equation for a known $\ga $ gives in a 
generic case  $0 < \vert \ga \vert < 2$ two possible 
pairs forming a rhombus, namely $(\al, \beta)$ and 
$(-\beta, -\al)$, so that for $\beta \ne \pm \al$ 
there is a single random variable for 
the two pairs:

%%%%%%%%%%%%%%%
\be
\label{susy.eq-rhombus}
 V_{\al ,\beta} = V_{-\beta ,-\al}
\ee
%%%%%%%%%%%%%%%

\noindent In the degenerate case $\ga =0$, it gives $2N$ equal pairs $V_0= 
V_{\al, \al}$, and in the degenerate cases $\vert \ga \vert = 2$ it gives a 
single pair $V_{\al, -\al}$. If, as explained in the introduction, the almost
degenerate rhombuses are neglected, which should be harmless since they are not 
generic, as well as the possible effect of nearest neighbor slight overlaps 
between the momentum definition of cells which may lead to slight complications, 
$V$ is well approximated by a random Gaussian $2N \times 2N$ complex matrix. 
This matrix is Hermitian because of (\ref{susy.eq-hermitian}):

%%%%%%%%%%%%%%%
\be  \label{susy.hermi}
V_{\al ,\beta} = \bar V_{\beta ,\al}
\ee
%%%%%%%%%%%%%%%

\noindent It is therefore convenient to label sectors as $\{1,\cdots,N\} \cup 
\{-N, \cdots,  -1\}$, so that $\{1,\cdots,N\}$ labels 
projective sectors, as shown in 
Figure 1.

%%%%%%%%%%%%%%%%%%%%%%%%%%%%%%%%%%%%%%%%%%%%%%%%%%%%%%%%%%%%%%%%%%%%%%%%%%%
\medskip
\centerline{\psfig{figure=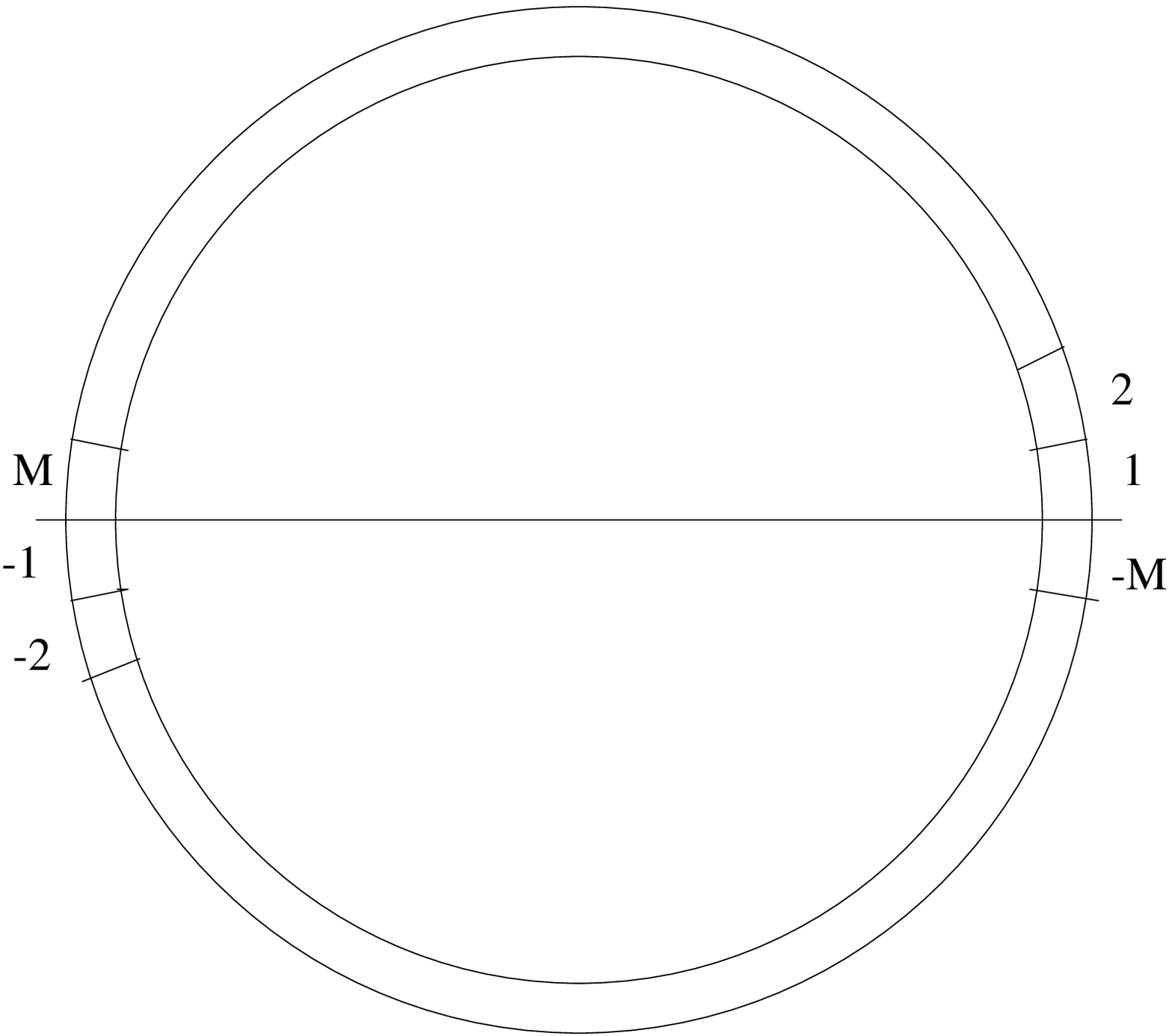,width=8cm}}
\medskip
\centerline{Figure 1. Numbering sectors}
\medskip
%%%%%%%%%%%%%%%%%%%%%%%%%%%%%%%%%%%%%%%%%%%%%%%%%%%%%%%%%%%%%%%%%%%%%%%%%%%

\noindent This leads to a set of 
$ 2 N(N-1)/2 = N(N-1)$ complex variables 
$ V_{\al,\beta} $ for $\al \in \{1,\cdots,N\}$, 
and $\beta \in \{1,\cdots,N\}\cup 
\{-N, \cdots,  -1\}, \al < \vert \beta \vert$. 
They fill a quarter of the matrix, made of the upper 
half of the upper-left quarter and of the lower half
of the upper-right quarter.
All other coefficients are deduced from these.
Thanks to (\ref{susy.eq-rhombus})-(\ref{susy.hermi}), and
to the particular numbering of Figure 1, 
there is symmetry around the antidiagonal, and conjugate
symmetry around the ordinary diagonal (so that $V$
is hermitian).

\noindent The upper anti-diagonal is made of $N$ additional independent
complex variables  $V_{\al, -\al } $ for $\al \in \{1,\cdots,N\}$, corresponding 
to momentum transfers of length $2$ of $V$, while the lower antidiagonal is 
hermitian conjugate to the upper one. Finally the variables $V_{\al \al}$ are 
all equal to a single real variable $V_0$ and that fills the ordinary diagonal 
of the matrix $V$.

\noindent The total number of real independent variables among ther matrix 
elements, is therefore $2N(N-1) + 2N + 1 = 1+2N^2$. They all have covariance 
equal to 1. The corresponding normalized Gaussian measure will be denoted $d\mu 
(V)$. This model is similar to the standard GUE (that would have $4N^2$ 
independent real variables for a $2N \times 2N$ matrix instead), except for the 
change on the diagonals and the additional symmetry with respect to the 
anti-diagonal, corresponding to rhombus flips. 

 The structure of $V$ can be summarized as follows
(remember that the rows and columns indices are labelled
as $\{1,\cdots,N,-N, \cdots,  -1\}$ hence {\it not}
as $\{1,\cdots,N,-1, \cdots,  -N\}$):

{\tiny
$$
\mbox{\normalsize $V$}\;=\;
 \left(
   \begin{array}{cc}
    \left[
     \begin{array}{ccccccc}
  V_0 &        & \vdots &        & \vdots       &        &  \\
      & \ddots & \vdots &        & \vdots       &        &  \\
      & \cdots &   V_0  & \cdots & V_{\al,\beta}& \cdots &  \\
      &        & \vdots & \ddots & \vdots       &        &  \\
   &\cdots&\overline{V_{\al,\beta}}&\cdots& V_0 & \cdots &  \\
      &        & \vdots &        & \vdots       & \ddots &  \\
      &        & \vdots &        & \vdots       &        & V_0  
 \end{array} 
     \right] & 
    \left[
     \begin{array}{ccccccc}
      &        & \vdots &        & \vdots       &        &  \\
      & \ddots & \vdots &        & \vdots       &        &  \\
      & \cdots &V_{\al,-\beta}& \cdots & V_{\al,-\al}& \cdots &  \\
      &        & \vdots & \ddots & \vdots       &        &  \\
      &\cdots&V_{\beta,-\beta}&\cdots&V_{\al,-\beta}&\cdots &  \\
      &        & \vdots &        & \vdots       & \ddots &  \\
      &        & \vdots &        & \vdots       &        &  
 \end{array} 
     \right] \\
--------------------& -------------------- \\
     
    \left[
     \begin{array}{ccccccc}
      &        & \vdots &        & \vdots       &        &  \\
      & \ddots & \vdots &        & \vdots       &        &  \\
& \cdots &\overline{V_{\al,-\beta}}& \cdots 
&\overline{V_{\beta,-\beta}}& \cdots&\\
      &        & \vdots & \ddots & \vdots       &        &  \\
&\cdots&\overline{V_{\al,-\al}}&\cdots&
\overline{V_{\al,-\beta}}&\cdots&\\
      &        & \vdots &        & \vdots       & \ddots &  \\
      &        & \vdots &        & \vdots       &        &  
 \end{array} 
     \right] & 
    \left[
     \begin{array}{ccccccc}
  V_0 &        & \vdots &        & \vdots       &        &  \\
      & \ddots & \vdots &        & \vdots       &        &  \\
      & \cdots &   V_0  & \cdots & {V_{\al,\beta}}& \cdots &  \\
      &        & \vdots & \ddots & \vdots       &        &  \\
      &\cdots&\overline{V_{\al,\beta}}&\cdots& V_0 & \cdots &  \\
      &        & \vdots &        & \vdots       & \ddots &  \\
      &        & \vdots &        & \vdots       &        & V_0  
 \end{array} 
     \right]
   \end{array}
 \right)     
$$
}%%fin.de\tiny

Our main result in this note is:

%%%%%%%%%%%%%%%
\begin{theorem}
In the large $N$-limit, the DOS of this Flip Matrix Model converges to Wigner's 
semi-circle distribution. The corrections to the limit are uniformly bounded as 
$O(1/N)$ as $N \to \infty$.
\end{theorem}
%%%%%%%%%%%%%%%

\noindent The rest of the paper is devoted to the proof of this result. 
Heuristically, the diagonals should not matter much for the average spectral 
properties. The main difficulty is to check that the additional symmetry of the 
rhombus flip does not alter Wigner's semi-circle law. This is made precise using 
the supersymmetric formalism \cite{Efet,Mir}.

%%%%%%%%%%%%%%%%%%%%%%%%%%%%%%%%%%%%%%%%%%%%%%%%%%%%%%%%%%%%%%%%%%%%%%%%%%%
\section{Mean field analysis}
\label{susy.sec-susy}

\noindent Let the superfields be

%%%%%%%%%%%%%%%
$$ 
\Psi_{\si \al} \;=\;
 (S_{\si \al} , \ch_{\si \al})
$$
%%%%%%%%%%%%%%%

\noindent for $\si = \pm 1$ and greek variables such as $\al$, $\beta$ now 
running only within $[1,...,N]$. The conventions are taken from \cite{Mir}. 
Hence bosonic fields are complex with $\bar S = S^{+}$ and pairs of 
anticommuting variables $\chi, \chi^{+}$ satisfy:
%%%%%%%%%%%%%%%
$$ 
\chi_i \chi_j \;=\;- \chi_j \chi_i \,, \;
 \chi_i^+ \chi_j \;=\; - \chi_j \chi_i^+ \,,\;
  \chi_i^+ \chi_j^+ \;=\; - \chi_j^+ \chi_i^+
$$
%%%%%%%%%%%%%%%

\noindent
%%%%%%%%%%%%%%%
$$ 
(\chi_i^+)^+ \;=\; - \chi_i \,,\;\;\;
 (\chi_i \chi_j)^+ \;=\; \chi_i^+ \chi_j^+ \,,
$$
%%%%%%%%%%%%%%%

\noindent

%%%%%%%%%%%%%%%
$$ 
\int d \chi_i  \;=\;  \int d \chi_i^+ = 0, 
 \qquad 
  \int \chi_i  d \chi_i  \;=\; 
   \int \chi_i^+ d \chi_i^+ \;=\;
    \frac{1}{\sqrt{2 \pi}}
$$
%%%%%%%%%%%%%%%

\noindent so that

%%%%%%%%%%%%%%%
$$ 
\int \prod_i  d \chi_i^+ d \chi_i e^{- \chi^+ M \chi} \;=\;
 \det (M/2\pi)
$$
%%%%%%%%%%%%%%%

\noindent whereas ordinary complex bosonic variables give

%%%%%%%%%%%%%%%
$$ 
\int \prod_i  d S_i^+ d S_i e^{ - S^+ M S} \;=\;
 \det{(M/2\pi)}^{-1}
$$
%%%%%%%%%%%%%%%

\noindent (and $M$ has to have a positive real part). By the usual rules of 
anticommuting integrals each Gaussian term is written as an integral over 
superfields. Hence the density of states can be written as

%%%%%%%%%%%%%%%
\be
\label{susy.eq-supsersym1}
\nu (E) \;=\;
 \lim_{Im E \to 0^+} 
  \frac{1}{\pi}
   \Im  \int S^{+}_{\al} S_{\al}  e^{\;\imath L_0} \;
    \prod_{\al , \si } 
     d \Psi_{\si \al}^{+}\; d \Psi_{\si \al}\; d\mu(V)
\ee
%%%%%%%%%%%%%%%

\noindent where the supersymmetric action decomposes as

%%%%%%%%%%%%%%%
$$ 
L_0 \;=\; (A_0 + B_0 + C_0)
$$
%%%%%%%%%%%%%%%

\noindent with

%%%%%%%%%%%%%%%
$$ 
A_0 \;=\;
 \sum_{\al < \beta , \si} 
  \left\{ \la V_{\al , \si\beta} \;
    \left(\Psi_{\al}^{+} \Psi_{\si \beta} +
   \Psi_{-\si\beta}^{+} \Psi_{- \al} 
     \right) + h.c. 
  \right\}
$$
%%%%%%%%%%%%%%%
%%%%%%%%%%%%%%%
$$
B_0 \;=\;
 \sum_{\al} \la V_{\al ,- \al} 
  \left(
    \Psi_{ \al}^{+} \Psi_{- \al} + h.c.
  \right)
$$
%%%%%%%%%%%%%%%
%%%%%%%%%%%%%%%
$$
C_0\;=\;
 E + \la V_0
  \sum_{\al, \si }
   \Psi_{\si \al}^{+} \Psi_{\si \al}
$$
%%%%%%%%%%%%%%%

\noindent correspond respectively to the main part, the anti-diagonal and the 
diagonal of the matrix. The Gaussian integration over the $V$ variables will be 
performed except for $V_0$ leading to the quartic terms
 
%%%%%%%%%%%%%%%
$$
\nu (E) = \lim_{Im E \to 0^+} \frac{1}{\pi}
Im  \int S^{+}_{\al} S_{\al}  e^{-L}  
\prod_{\al , \si } d \Psi_{\si \al}^{+} d \Psi_{\si \al} d\mu_0 (V_0) 
$$
%%%%%%%%%%%%%%%
%%%%%%%%%%%%%%%
$$
L = \la^2 ( A + B + C)
$$
%%%%%%%%%%%%%%%
%%%%%%%%%%%%%%%
\be
\label{susy.eq-termA}
A =  \sum_{\al < \beta , \si}  ( \Psi^{+}_{\al} \Psi_{\si \beta }  +
\Psi^{+}_{- \si\beta }\Psi_{-\al } )
(\Psi^{+}_{\si \beta }\Psi_{\al } + \Psi^{+}_{ -\al }\Psi_{-\si \beta })
\ee
%%%%%%%%%%%%%%%
%%%%%%%%%%%%%%%
$$  B = \sum_{\al }  \Psi^{+}_{\al }\Psi_{-\al }  \Psi^{+}_{-\al }\Psi_{\al }
$$
%%%%%%%%%%%%%%%
%%%%%%%%%%%%%%%
$$  C = (\la^{-1} E +  V_0 )
\sum_{\al, \si }  \Psi_{\si \al}^{+} \Psi_{\si \al}\ .
$$
%%%%%%%%%%%%%%%

\noindent The sum in (\ref{susy.eq-termA}) can be written as
%%%%%%%%%%%%%%%
$$
A \;=\;
 \sum_\si \Phi^+_{\al \beta;\,\si}\Phi_{\al \beta;\,\si}
  \hspace{.5cm} with \hspace{.5cm}
\Phi_{\al \beta;\,\si} \;=\;
(\Psi^{+}_{\si \beta }\Psi_{\al } + \Psi^{+}_{ -\al }
\Psi_{-\si \beta })\ .
$$
%%%%%%%%%%%%%%%

\noindent Exchanging $\al$ and $\beta$ in $A= \sum_\si \Phi^+_{\al \beta; 
\si}\Phi_{\al \beta; \si}$, both for $\si =\pm 1$ does not change the sum. This 
is because the bosonic part in $\Phi$ commute and the rule for the fermionic 
part leads to $\Phi^+_{\al \beta; +}= \Phi_{\beta \al; +}$, whereas for $\si 
=-1$, $\Phi_{\beta \al; -}= \Phi_{\al \beta, -}$. Hence

%%%%%%%%%%%%%%%
$$
A \;=\; 
 \frac{1}{2}
  \sum_{\al ,\beta , \si}  ( \Psi^{+}_{\al} \Psi_{\si \beta }  +
\Psi^{+}_{- \si\beta }\Psi_{-\al } )
(\Psi^{+}_{\si \beta }\Psi_{\al } + \Psi^{+}_{ -\al }\Psi_{-\si \beta })
+ D_{++} + D_{+-} +D_-\ ,
$$
%%%%%%%%%%%%%%%

\noindent where

%%%%%%%%%%%%%%%
$$ D_- = - (1/2) \sum_{\al } ( \Psi^{+}_{\al} \Psi_{- \al }  +
\Psi^{+}_{\al  }\Psi_{-\al } ) (\Psi^{+}_{- \al }\Psi_{\al }
+ \Psi^{+}_{ -\al }\Psi_{ \al }) = -2B
$$
%%%%%%%%%%%%%%%
%%%%%%%%%%%%%%%
$$ 
D_{++} = - (1/2) \sum_{\al, \si } ( \Psi^{+}_{\si \al} \Psi_{\si \al })^2
\hspace{1cm}
D_{+-}= - \sum_{\al} ( \Psi^{+}_{\al} \Psi_{\al }
\Psi^{+}_{- \al} \Psi_{- \al })
$$
%%%%%%%%%%%%%%%
%%%%%%%%%%%%%%%
$$  D_{+-} = - \sum_{\al }( S_{\al}^{+}S_{\al} +  \ch_{\al}^{+}\ch_{\al}  )
(S_{ -\al}^{+}S_{-\al} +  \ch_{-\al}^{+}\ch_{-\al})\ .
$$
%%%%%%%%%%%%%%%

\noindent The decomposition of $D_{+-}$ into its boson and fermion parts and 
adding $B$ above, eventually leads to $ L = \la^2 ( \cA + \cB + \cC )$ with

%%%%%%%%%%%%%%%
$$  \cA  =  (1/2)
\sum_{\al , \beta , \si}  ( \Psi^{+}_{\al} \Psi_{\si \beta }  +
\Psi^{+}_{- \si\beta }\Psi_{-\al } )
(\Psi^{+}_{\si \beta }\Psi_{\al } + \Psi^{+}_{ -\al }\Psi_{-\si \beta })
$$
%%%%%%%%%%%%%%%
%%%%%%%%%%%%%%%
$$  \cB = \sum_{\al} \cB_{\al}
$$
%%%%%%%%%%%%%%%
%%%%%%%%%%%%%%%
$$  \cB_{\al} = - 2 S^+_{\al }S_{-\al}S^+_{-\al }S_{\al}
- \frac{1}{2}\sum_{\si } ( \Psi^{+}_{\si \al} \Psi_{\si \al })^2
- ( \sum_{\si} S_{\si\al}^{+} \ch_{-\si\al}^{+} )
( \sum_{\si} S_{\si\al} \ch_{-\si\al})
$$
%%%%%%%%%%%%%%%
%%%%%%%%%%%%%%%
\be
\label{susy.eq-calcdef}
\cC = C = (\la^{-1}E+V_0) \sum_{\al, \si } 
\Psi_{\si \al}^{+} \Psi_{\si \al} . 
\ee
%%%%%%%%%%%%%%%

\noindent The term $\cB$ being a sum of {\em diagonal quartic} terms, will be 
neglected for a while. For indeed it cannot be written as the {\em square} of a 
sum over $\al$. However, the bosonic part of this term has the {\em wrong sign}, 
which may create some difficulties when performing the integration in 
(\ref{susy.eq-supsersym1}). This problem will be 
addressed in section VI.
The other terms, however, can be reorganized so as to get the square of a sum 
over the $\alpha$'s by pushing the terms with index $\alpha$ on the left and the 
ones with index $\beta$ on the right. This is possible thanks to the commutation 
rules for bosons and for fermions. The calculation is tedious but 
straightforward and gives: 
 
%%%%%%%%%%%%%%%
\bqa \cA &=& -(1/2) (\sum_{\al, \si} S^{+}_{\si\al} S_{\si\al})^2 
+ (1/2) (\sum_{\al, \si} \ch^{+}_{\si\al} \ch_{\si\al}  )(h.c.) \\ 
&&
 - 2 (\sum_{\al} S^{+}_{\al} S^+_{-\al} ) (h.c.)
  -(\sum_{\si \al} S_{\si \al} \ch^+_{\si \al} ) (h.c.)
   -(\sum_{\si \al} S^+_{\si \al} \ch^+_{-\si \al} ) 
   (h.c.) \ . \no
\eqa 
%%%%%%%%%%%%%%%

\noindent The squares can be {\em unfolded} by mean of an integration over 
auxiliary gaussian fields. This amount to introduce two real fields $a_0$ and 
$b_0$, one complex field $a, \bar a$ and two pairs of fermionic fields $\xi, 
\bar \xi$ and $\eta, \bar \eta$ with Gaussian measure

%%%%%%%%%%%%%%%
\bqa 
d\mu (a_0, b_0, a, \bar a, \bar \xi , \xi, \bar \eta,\eta) 
 &=&  
  e^{-V_0^2/2 -a_0^2/2 -b_0^2/2 -\vert a\vert^2/2-\xi^* \xi -\eta^* \eta}
   \times \cdots \no\\
& \cdots & 
 \frac{dV_0\,da_0\, db_0\, d^2a}{(2\pi)^{5/2}}\,
  d\xi^* d\xi d\eta^* d\eta\ .
\eqa
%%%%%%%%%%%%%%%

\noindent Therefore 

%%%%%%%%%%%%%%%
$$  e^{\la^2 (\cA + \cC)} =
\int d\mu \;\; e^{\,i \la \sum_{\al}  \Ph^+_{\al} R \Ph_{\al}}
$$
%%%%%%%%%%%%%%%

\noindent with 

%%%%%%%%%%%%%%%
\bqa
\sum_{\al}  \Ph^+_{\al} R \Ph_{\al} & = &
 (a_0 + V_0 + \la^{-1} E )
  \sum_{\al, \si} S^{+}_{\si\al} S_{\si\al} \no \\
  &+&
   (ib_0 + V_0 + \la^{-1} E )
    \sum_{\al, \si} \ch^{+}_{\si\al} \ch_{\si\al} \no\\
 &+&  \left(
     \bar a \sum_{\al} S_{\al} S_{-\al} + h.c.
    \right)  \\
 &+&   \left(
     \xi^* \sum_{\si \al} S^+_{\si \al} \ch_{\si \al}+ h.c.
    \right)
 +  \left(
     \eta^* \sum_{\si \al} S_{\si \al} \ch_{-\si \al} + h.c.
    \right) \no
\eqa
%%%%%%%%%%%%%%%

\noindent where $\Phi$ is the superfield $\Phi_{\alpha}^+ =  (S^+_{\al}, 
S_{-\al}, \ch^+_{\al}, \imath\ch_{-\al})$ and $R$ is a $4 \times 4$ supermatrix. 
Setting $A_0 =(a_0 + V_0 + \la^{-1} E)$ and $\imath B_0 = (\imath b_0 + V_0 + 
\la^{-1} E)$ this matrix is given by

%%%%%%%%%%%%%%%
\be
\label{susy.eq-rmatrix}
R\;=\;\left(
      \ba{cccc}
      A_0    &  a    & \xi^*  & -\imath\eta \\
      \bar a & A_0   & \eta^* & -\imath\xi  \\ 
        \xi  & \eta  & \imath B_0  & 0    \\ 
      -\imath\eta^* & -\imath \xi^* &   0    & -\imath B_0\\
      \ea 
    \right)
    \;=\;
      \left(
      \ba{cc}
       A    & \rho^{\ast} \\
       \rho & B
      \ea
      \right)  \ .
\ee
%%%%%%%%%%%%%%%

\noindent In order that the integration over the primitive fields be given in 
term of $\Phi$ it will be convenient to introduce the new fermionic fields
 
%%%%%%%%%%%%%%%
$$
\tilde\ch_{-\al}\;=\; \imath \ch^+_{-\al}\,,\;\;
 \tilde\ch^+_{-\al}\;=\; \imath \ch_{-\al}\,,
 \hspace{.5cm}
  d\tilde\ch_{- \al}\;=\; -\imath d\ch^+_{- \al}\,, \;\;
   d \tilde \ch^+_{- \al}\;=\; -\imath d\ch_{- \al}\, .
$$
%%%%%%%%%%%%%%%

\noindent This leads to the formula

%%%%%%%%%%%%%%%
$$
 \prod_{\al} 
  \int d^2 S_{\al} d^2 S_{-\al}
   d\ch^+_{\al} d\ch_{\al} 
    d\tilde\ch_{-\al}^+ d\tilde\ch_{-\al}
    e^{i \la \Ph^+_{\al} R \Ph_{\al}} 
    \;=\; 
     \left[ Sdet (\imath \la R) 
     \right]^{-N} \ .
$$
%%%%%%%%%%%%%%%

\noindent  The {\em super determinant} can be computed from the {\em Schur 
complement} formula

%%%%%%%%%%%%%%%
$$
Sdet 
  \left(
   \ba{cc} A& \rho^* \\
   \rho & B 
   \ea 
  \right )
  \;=\;
   \frac{1}{\det B}\; \det( A - \rho^* B^{-1} \rho )\ .
$$
%%%%%%%%%%%%%%%

\noindent Defining $\hat a_0 = a_0 + V_0 + \la^{-1} E $, $\hat b_0 = ib_0 + V_0 
+ \la^{-1} E $ and using the Pauli matrices $\sigma_2, \sigma_2, \sigma_3$ the 
superdeterminant of $R$ is given by gives

%%%%%%%%%%%%%%%
\bqa
Sdet R &=& 
 \frac{1}{( \hat b_0 )^2} \;
  \det \
   \biggl( 
     \hat a_0 {\bf 1} + a_1 \si_1 + a_2 \si_2   \\
     &&-\left(
        \ba{cc}
        \xi^*  & -\imath \eta \\
        \eta^* & -\imath \xi 
        \ea 
       \right)
        \left(
        \ba{cc}
         1 & 0 \\
         0 & 1 
        \ea 
        \right ) 
         \left(
         \ba{cc}
         \xi           &  \eta        \\
        -\imath \eta^* & -\imath \xi^*
         \ea
         \right )
           \hat b_0^{-1} 
   \biggr) \no\\
   &=& \frac{1}{( \hat b_0 )^2}
 \biggl( 
   \left\{
    (a_0 + V_0 + \la^{-1} E + ( \eta^* \eta + \xi^* \xi ) \hat b_0^{-1}  
   \right\}^2 \no \\
   &-& 
    \left\{ 
     a + 2 \xi^* \eta  \hat b_0^{-1}
    \right\}
    \left\{
     \bar a + 2 \eta^* \xi  \hat b_0^{-1}
    \right\}  
 \biggr) \ .\no
\eqa
%%%%%%%%%%%%%%%

\noindent Including back the scaling factors, taking into account the relation 
$2\la^2 = N^{-1}$ and ignoring the $\cB$ term, the density of states is given by

%%%%%%%%%%%%%%%
\bqa~
\label{susy.eq-dossdet}
 \int da_0 dV_0 &\cdots &
  \biggl(
    e^{- (a_0 ^2 + V_0^2 + \vert a \vert ^2 + b_0^2)}
    \frac{(E + V_0 + i b_0 )^2}{( E + V_0 + a_0 )^2-\vert a\vert ^2}
  \biggr)^N \nonumber\\
   &\cdots &\times 
    (1 + O(1/N))\;\;
     \frac{E + V_0 + a_0}{(E + V_0 + a_0 )^2  
     - \vert a\vert ^2} \ .
\eqa~
%%%%%%%%%%%%%%%

\noindent For indeed, taking into account the part of the $2$-point function 
involved in the integral (\ref{susy.eq-supsersym1}), the fermionic part gives a 
$O(1/N)$ correction in the form

%%%%%%%%%%%%%%%
$$
\langle (1+ r \eta^* \eta \xi^* \xi )^{-N-1}\rangle
 \;=\; 
  1 - r \frac{N+1}{4N^2}\,,
$$
%%%%%%%%%%%%%%%

\noindent thanks to the integral

%%%%%%%%%%%%%%%
$$
\frac{\pi}{2N} \int  d\xi^* d \xi
  e^{-2N \xi^* \xi } \xi^* \xi  \;=\;
   - \frac{1}{4N}\ .
$$
%%%%%%%%%%%%%%%

%%%%%%%%%%%%%%%%%%%%%%%%%%%%%%%%%%%%%%%%%%%%%%%%%%%%%%%%%%%%%%%%%%%%%%%%%%%
\section{The saddle points at leading order}
\label{susy.sec-saddlepoint}

\noindent In this section the $O(1/N)$ and $\cB$ terms are still neglected.
As $N\rightarrow \infty$, the integral (\ref{susy.eq-dossdet}) can be treated by 
the saddle point method, since no fermion variable are involved anymore. The 
saddle point equations can be written $\partial S =0$ where $S$ is the {\em 
effective action} given by 

%%%%%%%%%%%%%%%
$$
S = - (a_0^2  + V_0^2 + \vert a \vert ^2 + b_0^2)  +
2 \ln (E + V_0 + i b_0)
- \ln (( E + V_0 + a_0 )^2  - \vert a\vert ^2))\ .
$$
%%%%%%%%%%%%%%%

\noindent More precisely, differentiating with respect to the independant 
variables appearing in the integral (\ref{susy.eq-dossdet}) gives 

%%%%%%%%%%%%%%%
\be
\label{susy.eq-saddle1}
- (1/2) \frac{\partial S }{\partial a_0} \;=\;
 a_0 + \frac{ a_0 + V_0 + E}{(E + V_0 + a_0 )^2  
 - \vert a\vert ^2 } \;=\; 0\ ,
\ee
%%%%%%%%%%%%%%%
%%%%%%%%%%%%%%%
\be
\label{susy.eq-saddle2}
- (1/2) \frac{\partial S}{\partial \vert a \vert}
 \;=\; \vert a\vert [1 - \frac{1}{(E + V_0 + a_0 )^2  - \vert a\vert ^2 } ]
  \;=\;0 \ ,
\ee
%%%%%%%%%%%%%%%
%%%%%%%%%%%%%%%
\be 
\label{susy.eq-saddle3}
 -\frac{\imath}{2} \frac{\partial S}{\partial b_0} \;=\;
  ib_0 + \frac{1}{ E + V_0 + i b_0 } \;=\; 0\ ,
\ee
%%%%%%%%%%%%%%%
%%%%%%%%%%%%%%%
\be
\label{susy.eq-saddle4}
(1/2) \frac{\partial S}{\partial V_0}
= -V_0 + \frac{ 1}{E + V_0 + i b_0 }
- \frac{ a_0 + V_0 + E }{(E + V_0 + a_0 )^2  
- \vert a\vert ^2} = 0\ .
\ee
%%%%%%%%%%%%%%%

\noindent Thanks to (\ref{susy.eq-saddle1}) \& (\ref{susy.eq-saddle3}), 
(\ref{susy.eq-saddle4}) can be written as

%%%%%%%%%%%%%%%
\be
\label{susy.eq-saddle4i}
V_0 +  i b_0  - a_0 =  0\,,
\ee
%%%%%%%%%%%%%%%

\noindent so that, thanks to (\ref{susy.eq-saddle3}),

%%%%%%%%%%%%%%%
\be
\label{susy.eq-saddle3i}
i b_0 = \frac{-1}{E+a_0}\,.
\ee
%%%%%%%%%%%%%%%

\noindent Then two cases occur:

\begin{itemize}
 \item[\bf Case 1~:]$\vert a \vert = 0$. Substituting in (\ref{susy.eq-saddle1}) 
leads to
 %%%%%%%%%%%%%%%
 $$  (E+ 2a_0) (a_0^2 + a_0 E + 1) = 0\,,
 $$
 %%%%%%%%%%%%%%%
 giving two solutions
 
   %%%%%%%%%%%%%%%
   \be
   \label{susy.eq-solusadlle1}
   a_0=
    \frac{-E \pm i \sqrt{4 - E^2}}{2}\,,\;\;
      \imath b_0=
       \frac{ -E \pm i \sqrt{4 - E^2}  }{2}\,,\;\;
         V_0= 0\ ,
   \ee
   %%%%%%%%%%%%%%%
   %%%%%%%%%%%%%%%
   \be
   \label{susy.eq-solusadlle2}
   a_0\;=\;\frac{ -E}{2}\,,
    \hspace{.5cm}
     \imath b_0 \;=\;  \frac{ - 2}{E}\,,
      \hspace{.5cm}
       V_0\;=\; \frac{4 - E^2}{2E}\,.
   \ee
%%%%%%%%%%%%%%%
 \item[\bf Case 2~:]$\vert a \vert \neq 0$. Then, (\ref{susy.eq-saddle1}), 
(\ref{susy.eq-saddle2}) \& (\ref{susy.eq-saddle4i}) give
 
  %%%%%%%%%%%%%%%
  $$
   a_0 = -\frac{V_0+E}{2}\,,\;\;
    |a|^2 = (\frac{V_0+E}{2})^2 -1\,,\;
     \imath b_0 = -\frac{3V_0+E}{2}\ ,
  $$
  %%%%%%%%%%%%%%%
  %%%%%%%%%%%%%%%
  $$   
  3V_0^2 -2EV_0 + 4 -E^2 =0\ .
  $$
  %%%%%%%%%%%%%%%

\noindent It leads to the following solution
   %%%%%%%%%%%%%%%
   \bqa~
   \label{susy.eq-solusadlle3}
   V_0  &=& \frac{E \pm 2 i \sqrt{3 - E^2}  }{3} \ ,\no\\
   a_0 &=& -\frac{2E \pm  i \sqrt{3 - E^2}}{3} \ ,\no\\ 
   \imath b_0 &=& - E \mp  i \sqrt{3 - E^2} \ ,\no\\
   \vert a \vert ^2 &=& \frac{5E^2 -12  \pm  i 4E 
   \sqrt{3 - E^2}}{9}\ .
   \eqa~
%%%%%%%%%%%%%%%

\end{itemize}

\noindent  As $N \to \infty$ the leading contributions are given by the value of 
the action at the saddle, and we find, for the first saddle point
%%%%%%%%%%%%%%%
$$  S = 0  \ .
$$
%%%%%%%%%%%%%%%
For the second saddle point, if $E^2 < 4$,
%%%%%%%%%%%%%%%
$$
S = \frac{4 - E^2}{2} + 4 \ln (|E|/2) < 0\,,
$$
%%%%%%%%%%%%%%%
and, for the third saddle point, if  $E^2 \le 3$,
%%%%%%%%%%%%%%%
$$
S = E^2/3  - \ln 3 \pm 2i (-\phi + \sin 2 \phi)\,,
\hspace{.5cm}\Rightarrow \hspace{.5cm}
 \Re{S} <0\,,
$$
%%%%%%%%%%%%%%%
where $E/\sqrt 3 = \cos \phi$, $\sqrt{3 -E^2}/\sqrt 3 = \sin \phi$. 
Consequently, in the large $N$ limit, only the first saddle point contributes to 
the density of states, and gives the usual semi-circle law. Therefore our model
as expected is in the class of universality of the semi-circle law as $N \to 
\infty$. The saddle points are degenerate and one should bound the difference 
between saddle point contribution and the correction corresponding to 
out-of-the-saddle part of the integral. This should be done e.g. like in 
\cite{DPS}.

%%%%%%%%%%%%%%%%%%%%%%%%%%%%%%%%%%%%%%%%%%%%%%%%%%%%%%%%%%%%%%%%%%%%%%%%%%%
\section{The diagonal quartic terms}
\label{susy.sec-quartic}

\noindent In this section, the corrections due to the $\cB$ term are shown to be 
small as $N \to \infty$. As noticed in Section~\ref{susy.sec-susy}, the bosonic 
part of the quartic term $\cB$ has the wrong sign. For this reason before 
integrating over superfields it will be convenient to write a Taylor expansion 
with integral remainder successively for each of the $N$ sectors appearing in 
the sum for $\cB$: 

%%%%%%%%%%%%%%%
$$  
\cB = \sum_{\al} \cB_{\al}\ .
$$
%%%%%%%%%%%%%%%

\noindent To first order the expansion gives

%%%%%%%%%%%%%%%
$$
e^{-\cB_{\al}} = 1 - \int_{0}^{1}
\cB_{\al} e^{-\cB_{\al}} e^{+t\cB_{\al}} dt\ .
$$
%%%%%%%%%%%%%%%

\noindent This Taylor expansion either suppresses $\cB_{\al}$ from the 
exponential of the action or generates a remainder term $ R_{\al} =
- \int_{0}^{1}\cB_{\al} e^{-\cB_{\al}} e^{+t\cB_{\al}} dt$. Let $p(N)$ be the 
integer part of $N/\sqrt{\log N}$. The expansion is stopped to the order $p(N) = 
p$. This gives:

%%%%%%%%%%%%%%%
$$
e^{-\cB} =  1 +  
 \sum_{P \subset [1,...,N] \atop 0< |P| \le p(N)} R_P\,,
$$
%%%%%%%%%%%%%%%

\noindent where

%%%%%%%%%%%%%%%
$$ 
 R_P = \prod_{\al \in P} R_{\al} \quad  {\rm if} \  
 |P| < p(N)\ ,
$$
%%%%%%%%%%%%%%%
%%%%%%%%%%%%%%%
\be
\label{susy.eq-rpterm}
R_{P} = 
\prod_{\al \in P} R_{\al} \prod_{\al > \max P} e^{-\cB_{\al}} 
\quad {\rm if} \  |P| = p(N)\ .
\ee
%%%%%%%%%%%%%%%
Let $Q$ be the complement of $P$ in $[1,...,N]$. The term $1$ was treated in the 
previous sections. The remainders terms $R_P$ must be shown to be $O(1/N)$ as
$N \to \infty$. Let $R_{p(N) } = \sum_{P \subset [1,...,N] \atop
|P| = p(N)}R_{P}$ be considered first. For this term it is not necessary to 
perform any saddle point analysis and it is sufficient to return to the original 
expression given in (\ref{susy.eq-supsersym1}), hence to a functional integral
over $V$. However some $e^{-\cB_{\al}} $-factors are 
missing or appear with reduced 
weights. To correct for this, all the $e^{-\cB_{\al}}$-terms are recombined with 
the $\cA$-term to reproduce the initial functional integral 
(\ref{susy.eq-supsersym1}). However, there are quartic correction terms
$e^{t\cB_{\al}}$ or $e^{\cB_{\al}}$. The important remark is that the bosonic 
part of these terms has now the right sign~! Therefore they can be represented 
as a well defined integral over a new auxiliary field $W_{\al}$. For instance

%%%%%%%%%%%%%%%
$$ 
 e^{ -2t S^+_{\al }S_{-\al}S^+_{-\al }S_{\al}} = 
  e^{ -2t \vert S_{\al } S_{-\al }\vert ^2} = 
   \int d W_{\al} d \bar W_{\al}
    e^{- \vert W_{\al} \vert ^2} e^{\imath\sqrt{2t} W_{\al} S_{\al } 
    \bar S_{-\al } + cc}\ .
$$
%%%%%%%%%%%%%%%

\noindent With slightly condensed notations, this leads to 

%%%%%%%%%%%%%%%
$$ 
R_{p(N) }  = 
 \sum_{P \subset [1,...,N] \atop |P| = p(N)}  
  \int  S^+S 
  \biggl( 
   \prod_{\al \in P} 
    \int_{0}^{1}\cB_{\al} dt
  \biggr) 
     e^{i \Psi^+ (E - V + \sqrt{2t}W) \Psi}
      d \Psi^+ d\Psi d\mu (V, W)\ .
$$
%%%%%%%%%%%%%%%

\noindent Then a complex translation $V_0 \mapsto V_0 \pm\imath \la^{-1}$ is 
performed, with the same sign as the imaginary part of $E$ in order to avoid  
crossing of singularities. In other words

%%%%%%%%%%%%%%%
$$ \int_{-\infty}^{+\infty} e^{-V_0^{2}} F(V_0) dV_0 =
\int_{-\infty}^{+\infty} e^{-V_0^{2}}
e^{-2i\la^{-1}V_0^{2}} e^{\la^{-2}}  
F(V_0 + i \la^{-1}) dV_0\ .
$$
%%%%%%%%%%%%%%%

\noindent The functional integral can now be bounded by its absolute
values everywhere, namely the following contributions are bounded

- by $2^N$, for the sum over $P$, that is the total number of subsets of 
$[1,N]$;

- by $1$, for the integrals such as $\int_{0}^{1} dt$;

- by 1, for the oscillating imaginary integrals;

- by Gram's inequality for fermions or the Schwarz inequality for bosons, for 
the remainders terms $\cB_{\al}$;

- by 1, for every propagator since the imaginary translation in $V_0$
has created an imaginary part proportional to the identity in the denominator of 
the Green's function.

\vspace{.2cm}

\noindent This means that each $\cB_{\al}$ term give rise to a small factor
$\la^{2}=1/N$ for each sector in $P$, hence a factor $1/N^{p(N)}$.
The two source terms are bounded by 1. The normalization determinants are then 
easily bounded by $c^{N}$, even without using the supersymmetry
cancellations, since the operators considered are bounded in a finite
$2N$ dimensional space thanks to the imaginary part of $E$ which is
no longer infinitesimal. Combining all factors leads to

%%%%%%%%%%%%%%%
$$
R_{p(N) }  \le  
 c^{N}  e^{-c N \sqrt{\log N}}\,,
$$
%%%%%%%%%%%%%%%

\noindent showing that this correction term is small indeed.

\vspace{.2cm}

\noindent It remains to prove that the sum of the terms $R_P$ with $1\le \vert P 
\vert \le p(N)$ is small. Let $p=\vert P\vert$. To bound these terms a  
mean-field analysis will be performed (see Section~\ref{susy.sec-susy} \& 
\ref{susy.sec-saddlepoint}) in terms of the $a$ and $b$ fields, but only for the 
sectors of the theory in the complement $Q$ of $P$. The functional integral to 
be bounded for a single term is (\ref{susy.eq-rpterm}). Now, the 
$e^{-\cB_{\al}}$-terms are recombined only for $\al \in P$ with the $\cA$ term 
to reproduce the initial functional integrals over the $V$ fields and the terms 
$e^{t\cB_{\al}}$, also for $\al \in P$, are again given by defined integrals 
over new auxiliary fields $W_{\al}$. Finally, the quartic terms, with sector 
sums reduced to $Q$, are treated exactly as in the previous section, hence a 
mean-fields $a, a_0, b_0$  are correspondingly introduced. This leads to a 
representation

%%%%%%%%%%%%%%%
$$
R_P = 
 \int dV_{P,P}\, dV_{P,Q}\, d^2a \,dV_{0} \,da_0\, db_0
  \prod_{\al \in P} R_{\al}\; e^{\cL_{Q}}\ ,
$$
%%%%%%%%%%%%%%%

\noindent where $Q$ is the complement of $P$ in $[1,...,N]$. In addition, 
$V_{P,P}$ is the part of the matrix $V$ corresponding to rows and columns in 
$P$, including the new fields of the $W$ type. $V_{P,Q}$ correspond to one entry 
in $P$ and the other in $Q$, and the mean field computation is now retricted to 
$Q$. The integral over superfields gives rise again to a superdeterminant and an 
additional corrections, of the type

%%%%%%%%%%%%%%%
$$
\exp{[Tr\log (1+ C V)]}\ ,
$$
%%%%%%%%%%%%%%%

\noindent where $C= R ^{-1}$ is the matrix for the $N-p$ analogous problem, as 
in the previous Section~\ref{susy.sec-susy} (see eq.~(\ref{susy.eq-rmatrix})), 
and $V= V_{PP} + V_{P,Q}$ is the perturbation. This correction term is bounded 
by

%%%%%%%%%%%%%%%
$$ 
| \exp{[Tr\log (1+ K)]}  | \le  
\exp{(Tr K+K^{*} + KK^{*})}\ .
$$
%%%%%%%%%%%%%%%

\noindent Evaluating $C= R ^{-1}$ at the saddle point costs a factor 
$\exp{(N p/2N)} = e^{2p}$ at most. Each term $\cB_{\al}$ naively gives a factor 
$\la^{2}=1/(2N)$ when evaluated, but this simply compensates 
the sum over $\al$ when $p$ is nonzero but 
small, so we have to gain an additional $1/N$ factor. 
Adding a few expansion steps
gives such a small additional factor $1/N$ (already for
the first non zero value $p=1$). 
This can be seen by integrating by parts the 
superfields in the {\em vertex} $ \cB_{\al}$ which has been taken down the 
exponential. By supersymmetry, the {\em vaccum graph} 
corresponding to a contraction of the four 
fields at the vertex
vanishes. This is absolutely necessary since
this graph by simple scaling is proportional 
$1/N$ and cannot have any additional $1/N$ factor. 
Its vanishing can be checked by hand:

- the self-contractions of the bosonic piece $2 S^+_{\al }S_{-\al}S^+_{-\al 
}S_{\al}$ give a factor +2;

- the self-contractions of the boson-fermion piece $$ ( \sum_{\si} 
S_{\si\al}^{+} \ch_{-\si\al}^{+} ) ( \sum_{\si '} S_{\si\al} \ch_{-\si\al}) $$ 
give a factor -2 (since there is one fermionic loop giving the minus sign
and one sum over $\si$ giving a factor 2 only after the contractions);

- the selfcontractions of the term $\frac{1}{ 2}\sum_{\si } ( \Psi^{+}_{\si \al} 
\Psi_{\si \al })^2$ are clearly supersymmetric and also add up to 0.

\vspace{.2cm}

\noindent Consequently, at least one field of the 
{\em vertex} $ \cB_{\al}$
has to contract to the exponential. Performing two 
contractions in turn, gives 
always at least a factor $p/N^2$ at the end instead of the naive $1/N$ factor. 
Indeed the worst case corresponds to the non trivial contraction term being of
the type $V_{PQ}$. This generates a new factor 
$1/N$ but a new sum over $\beta 
\in Q$, which costs $N-p \simeq N$ so nothing is gained yet. 
But contracting this $\beta$ field again either 
generates a diagonal term, hence a new $1/N$ factor, 
with no new sum, or returns to a $V_{Q,P}$ 
term. This last situation generates 
a new $1/N$ factor and a new sum over 
sectors $\gamma$ but this time this new sum is restricted
to $P$, so it costs only a factor $p$ instead of $N$! 
Hence at worst, after these two contraction steps, 
a total factor $p/N^2$ instead of the naive factor
$1/N$ is associated to each 
vertex, as announced. Now the sum over $P$ costs a total factor 
$N!/p!(N-p)!$, hence is bounded by $c^{p} [N/p]^{p}$. Combining all factors, the 
sum of contributions of such terms with $p\ne 0$ is bounded by $ 
\sum_{p=1}^{+\infty}[c/N]^{p} \le c'/N$. It is therefore at least as small as 
$1/N$.

\vspace{.2cm}

\noindent This technique could also apply to the term $\cC$ in the formula
(\ref{susy.eq-calcdef}). Indeed its presence does not modify the large $N$ 
density of states. This remark might be used to simplify section 3. However it 
is quite natural to consider the diagonal part of $V$ on the same footing that 
the other mean field terms in the supermatrix $R$.

\medskip
\noindent{\bf Acknowledgments}
We thank A. Abdesselam for discussions on this work. J. Bellissard also
thanks H. Schulz-Baldes for discussions. He also thanks the Centre de
Physique Th{\'e}orique de l'{\'E}cole Polytechnique and the Institut des Hautes
{\'E}tudes Scientifiques for support during preparation of this work.
\medskip

%%%%%%%%%%%%%%%%%%%%%%%%%%%%%%%%%%%%%%%%%%%%%%%%%%%%%%%%%%%%%%%%%%%%%%%%%%%%%%%%
% \newpage

\end{document}